\documentclass[11pt]{article}
\usepackage{fontspec}
\usepackage[margin=0.96in]{geometry}
\usepackage{amsmath,amssymb,amsthm,mathtools}
\usepackage{booktabs,tabularx,array}
\usepackage{enumitem,microtype,xspace}
\usepackage{pifont}
\usepackage{xcolor,graphicx}
\usepackage[most]{tcolorbox}
\usepackage{algorithm,algpseudocode}
\usepackage{caption,subcaption,placeins,float,wrapfig}
\usepackage{url}
\usepackage{fancybox}
\usepackage{siunitx}
\usepackage{colortbl}
\definecolor{VoxDeep}{HTML}{213555}
\definecolor{VoxBlue}{HTML}{213555}
\definecolor{VoxPurple}{HTML}{6F42C1}
\definecolor{VoxLavender}{HTML}{F2EAFB}
\definecolor{VoxMid}{HTML}{AEB8C4}
\definecolor{VoxSoft}{HTML}{EBEEF2}
\definecolor{VoxMist}{HTML}{F8F9FA}
\definecolor{VoxEdge}{HTML}{D9DEE5}
\usepackage[colorlinks=true,linkcolor=black,citecolor=black,urlcolor=black]{hyperref}
\usepackage[nameinlink,capitalise,noabbrev]{cleveref}
\usepackage{natbib}
\newcommand{\VoxBestCell}[1]{\textcolor{VoxBlue}{\textbf{#1}}}

\newcolumntype{Y}{>{\raggedright\arraybackslash}X}
\newcolumntype{L}[1]{>{\raggedright\arraybackslash}p{#1}}

\setlist[itemize]{leftmargin=1.5em,itemsep=0.1em,topsep=0.25em}
\setlist[enumerate]{leftmargin=1.5em,itemsep=0.15em,topsep=0.25em}
\hypersetup{pdftitle={VoxMel: Wavelet-Factorized Autoregressive Mel-Spectrogram Modeling for Zero-Shot Speech Synthesis}}

\usepackage{titlesec}

\titleformat{\section}
  {\normalfont\Large\bfseries\color{VoxDeep}}
  {\thesection}{0.6em}{}

\titleformat{\subsection}
  {\normalfont\large\bfseries\color{VoxDeep}}
  {\thesubsection}{0.55em}{}

\titleformat{\subsubsection}
  {\normalfont\normalsize\bfseries\color{VoxDeep}}
  {\thesubsubsection}{0.5em}{}

\titleformat{\paragraph}[runin]
  {\normalfont\normalsize\bfseries\color{VoxDeep}}
  {}{0pt}{}

\titlespacing*{\section}{0pt}{1.8ex plus .5ex minus .2ex}{.8ex plus .2ex}
\titlespacing*{\subsection}{0pt}{1.3ex plus .3ex minus .1ex}{.5ex plus .1ex}
\titlespacing*{\paragraph}{0pt}{1.0ex plus .2ex}{0.6em}

\arrayrulecolor{black}
\begin{document}

\vspace*{-0.55in}
\noindent
\includegraphics[height=1.5cm]{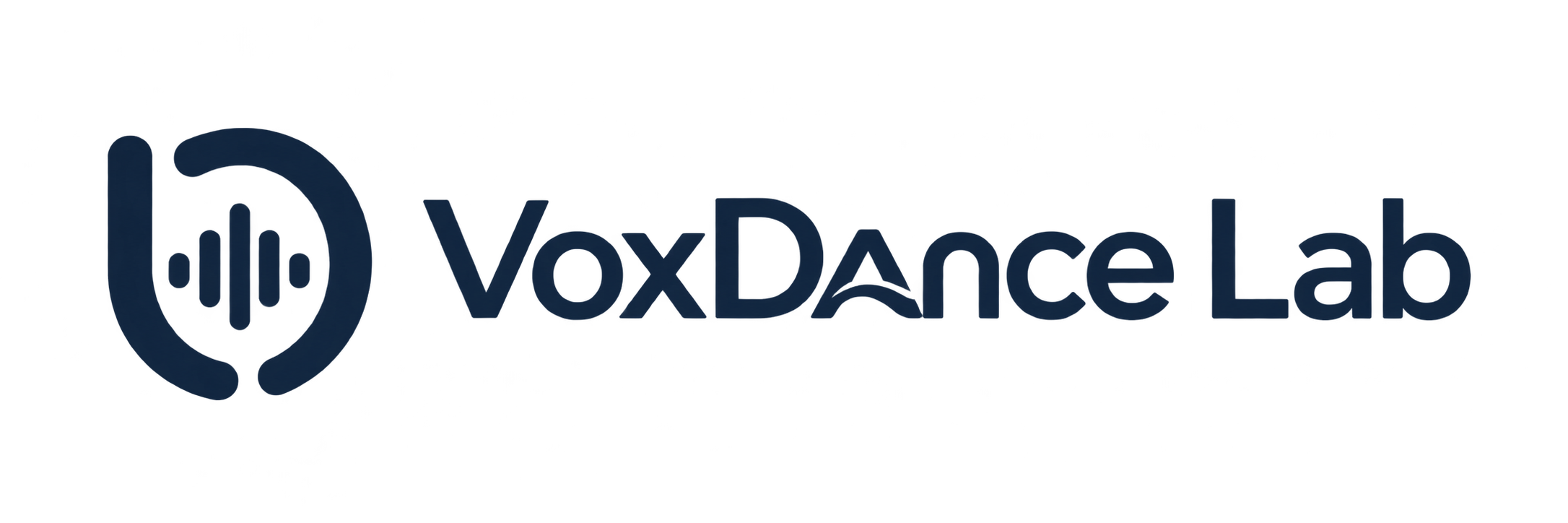}
\par\vspace{-1.5em}

\noindent{\color{VoxDeep}\rule{\textwidth}{2pt}}\par
\vspace{1.0em}
\begingroup
\centering
{\LARGE \rmfamily\bfseries\color{VoxDeep}
VoxMel TTS Technical Report\par}
\endgroup

\vspace{1.0em}

\noindent{\color{VoxDeep}\rule{\textwidth}{1pt}}\par

\vspace{1.0em}

\begingroup
\centering
% {\normalsize \bfseries\color{VoxDeep} \textbf{VoxDance Lab}\textsuperscript{1,2,3,4,5,6}\par}
{\normalsize \bfseries\color{VoxDeep} \textbf{Wenhao Guan}\textsuperscript{1,2}~~}
{\normalsize \bfseries\color{VoxDeep} \textbf{Jiaxin Ye}\textsuperscript{4}~~}
{\normalsize \bfseries\color{VoxDeep} \textbf{Haitao Li}\textsuperscript{5}~~}
{\normalsize \bfseries\color{VoxDeep} \textbf{Yifan Duan}\textsuperscript{3}~~}
{\normalsize \bfseries\color{VoxDeep} \textbf{Junxi Liu}\textsuperscript{3}~~}
{\normalsize \bfseries\color{VoxDeep} \textbf{Wangjin Zhou}\textsuperscript{6}\par} 
{\normalsize \bfseries\color{VoxDeep} \textbf{Lin Li}\textsuperscript{2}~~}
{\normalsize \bfseries\color{VoxDeep} \textbf{Qingyang Hong}\textsuperscript{2}~~}
{\normalsize \bfseries\color{VoxDeep} \textbf{Xie Chen}\textsuperscript{1,3}~~}

\vspace{0.35em}
{\footnotesize \color{VoxDeep}
\textsuperscript{1}Shanghai Innovation Institute~~\textsuperscript{2}Xiamen University~~\textsuperscript{3}Shanghai Jiao Tong University  \\[0.10em] 
\textsuperscript{4}Fudan University~~\textsuperscript{5}Zhejiang University~~\textsuperscript{6}Kyoto University  \\[0.10em]
\par}
\endgroup

\vspace{1.2em}
\begin{tcolorbox}[enhanced,width=\textwidth,colback=VoxMist,colframe=VoxEdge,boxrule=0.9pt,arc=10pt,left=14pt,right=14pt,top=10pt,bottom=10pt]
\begin{center}
{\color{VoxDeep}\bfseries Abstract}
\end{center}
\vspace{-0.25em}
Continuous autoregressive TTS relies on an acoustic target that enables stable next-latent modeling while preserving fine-grained acoustic details. While existing approaches employ learned compressed representations
(e.g., continuous VAE latents) to render prediction tractable, they rely on complex neural encoders that often cap generation fidelity. Conversely, training-free targets like mel-spectrograms eliminate encoder overhead, yet their potential for autoregressive generation remains largely underexplored. In this work, we propose
VoxMel, a tokenizer-free framework for autoregressive TTS that makes stable mel modeling tractable through a
fixed, exactly invertible two-axis Haar transform. The transform first decomposes each mel spectrogram along
frequency into local averages and a frequency residual, then splits the averaged component along time into
Coarse-Mel and a temporal residual. A language-model backbone with a flow-matching head autoregressively
generates Coarse-Mel. Non-autoregressive residual tiers then predict the temporal residual from Coarse-Mel,
followed by the frequency residual conditioned on the reconstructed Frequency-Averaged Mel. Neither residual is fed back into the autoregressive history.

Trained on 100K hours of Emilia, the supervised 0.6B model obtains
1.51\% WER and 0.738 speaker similarity (SIM) on test-en, and 1.29\%
CER and 0.755 SIM on test-zh. GRPO improves these results to 1.15\% WER
/ 0.753 SIM and 1.17\% CER / 0.776 SIM, respectively. We also distill
the guided flow head into two- and four-step students, requiring 2 or 4
head evaluations per coarse frame. The four-step student
retains competitive SIM and nearly
matches the supervised model in WER/CER.

\par
\vspace{0.6em}
\begin{flushleft}
{\small\textbf{Project page:} \href{https://voxdance.github.io/voxmel-demo}{https://voxdance.github.io/voxmel-demo}\par}
{\small\textbf{Github:} \href{https://github.com/VoxDance-Lab/VoxMel}{https://github.com/VoxDance-Lab/VoxMel}\par}
{\small\textbf{HuggingFace:} \href{https://huggingface.co/VoxDance-Lab/}{https://huggingface.co/VoxDance-Lab}\par}
\end{flushleft}
\end{tcolorbox}
\vspace{1.2em}

\section{Introduction}
Zero-shot text-to-speech (TTS) has advanced rapidly by casting speech
generation as autoregressive sequence modeling with language-model
backbones \citep{wang2023valle,du2024cosyvoice,anastassiou2024seed}.
A central design choice in these systems is the acoustic representation
to be generated. Most existing approaches autoregress over learned,
compressed representations, including discrete codec tokens
\citep{defossez2022encodec,kumar2023dac,wang2023valle} and continuous
latents from trained autoencoders
\citep{peng2025vibevoice,zhou2025voxcpm}. Consequently, synthesis
quality depends in part on the learned encoder and the fidelity of its
reconstruction.

Mel spectrograms offer a complementary alternative. As a standard
acoustic target in TTS \citep{shen2018tacotron2,ren2020fastspeech2},
they avoid the need for a learned acoustic tokenizer and remain
interpretable and model-independent. Although a vocoder is still
required to convert mel spectrograms into waveforms, it can be
independently upgraded without retraining the autoregressive generator.
Despite these advantages, mel spectrograms have received limited
attention as continuous autoregressive targets \citep{meng2024melle}.
Direct autoregression over full-band mel spectrograms presents two
challenges. Strong temporal redundancy may let a model rely on acoustic
history under teacher forcing instead of using the text condition.
Moreover, each sampled frame, including its fine-grained details, is fed
back into the history, where prediction errors may accumulate. Frame patching
\citep{jia2025ditar} and low-frame-rate learned latents
\citep{peng2025vibevoice,zhou2025voxcpm} reduce the number of
autoregressive steps, but their generated patches or latents still enter
the history. Token-wise coarse-to-fine methods
\citep{wang2025felle,wu2025clear} likewise feed reconstructed full-band
frames into later steps.

We propose \textbf{VoxMel}, a tokenizer-free TTS framework that applies
a fixed, exactly invertible two-axis Haar transform to mel spectrograms.
The transform first splits the frequency axis into Frequency-Averaged
Mel and a frequency residual, then splits Frequency-Averaged Mel along
time into a compact Coarse-Mel sequence and a temporal residual.

A language-model backbone with a flow-matching head generates
Coarse-Mel autoregressively. Two non-autoregressive tiers then predict
the temporal residual from Coarse-Mel and the frequency residual from
the reconstructed Frequency-Averaged Mel. Because neither residual
enters the autoregressive history, the model generates a compact coarse
sequence while retaining enough information to reconstruct the complete
mel spectrogram. We evaluate VoxMel on LibriSpeech and Seed-TTS-eval,
including ablations of residual feedback, comparisons with other speech
generation systems, and experiments on GRPO post-training and flow-head
distillation.

Our contributions are:
\begin{itemize}
  \item \textbf{Wavelet-factorized mel modeling.}
  VoxMel applies a fixed, exactly invertible two-axis Haar wavelet
  transform to mel spectrograms. A language-model backbone autoregresses
  only over Coarse-Mel, while non-autoregressive residual tiers predict
  the temporal and frequency residuals in dependency order.

  \item \textbf{Post-training and distillation.}
  Frozen-head GRPO improves all four Seed-TTS-eval metrics. We also
  distill the guided flow head into guidance-conditioned students with
  two or four inference steps, reducing head evaluations per coarse
  frame from 96 to 2 or 4. The four-step student retains competitive SIM and nearly matches the supervised
  model in WER/CER.

  \item \textbf{Evaluation and analysis.}
  We evaluate zero-shot synthesis on LibriSpeech test-clean and
  Seed-TTS-eval test-en/test-zh.
  Decomposition ablations examine how removing residuals from the
  autoregressive history affects text reliance and end-to-end synthesis.

\end{itemize}

% ======================================================================
\section{Preliminaries}
\label{sec:prelim}

Let $X = (x_1, \dots, x_T) \in \mathbb{R}^{T \times M}$ denote a log-mel spectrogram with $T$ frames and $M$ mel channels, and let $y$ denote the
conditioning information (text and a speaker prompt). Continuous
autoregressive TTS factorizes the conditional density as
\begin{equation}
  p(X \mid y) \;=\; \prod_{t=1}^{T} p\!\left(x_t \mid x_{<t},\, y\right),
  \label{eq:ar}
\end{equation}
where a language-model backbone summarizes the prefix in a hidden state
$h_t = f_\theta(x_{<t}, y)$ and a sampling-based prediction head draws
$x_t \sim p_\phi(\cdot \mid h_t)$. The head may be a variational latent
head \citep{meng2024melle}, a diffusion head
\citep{jia2025ditar}, or a conditional flow-matching head that
transports Gaussian noise to the target conditioned on $h_t$
\citep{lipman2022flow}. Patch-based variants \cite{jia2025ditar,zhou2025voxcpm}
group $P$ consecutive frames into one autoregressive step, with the head
modeling the joint distribution of a $P \times M$ block.
Training uses teacher forcing. At inference, each sampled frame or patch
is fed back into the history.

% ======================================================================
\section{VoxMel}
\label{sec:method}

\begin{figure*}[t]
  \centering
  \includegraphics[width=1.0\linewidth]{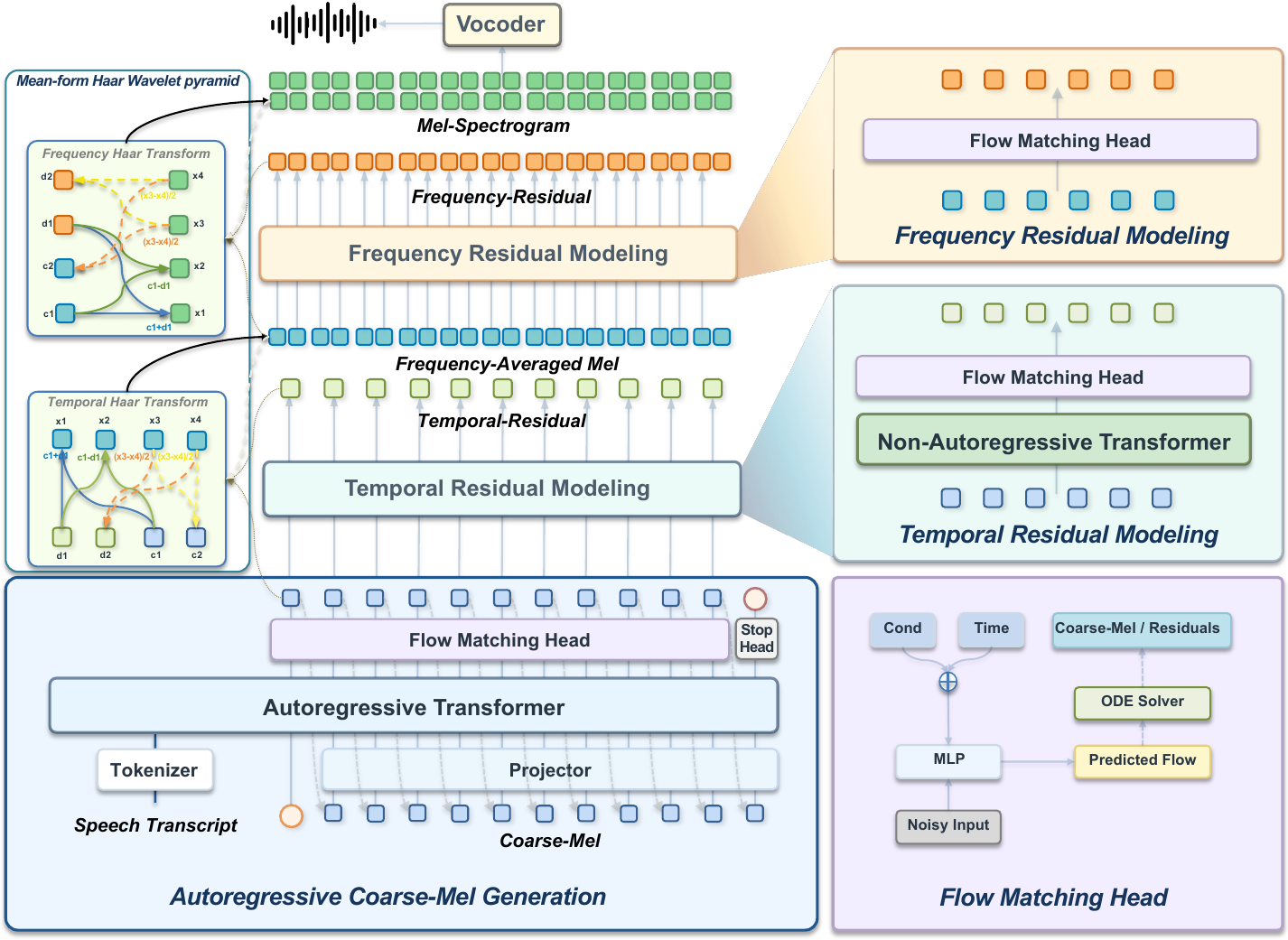}
\caption{\textbf{Overview of VoxMel}.
\emph{Top left}: Dashed arrows show the two-axis Haar decomposition.
The mel spectrogram is first split along frequency into
Frequency-Averaged Mel and a frequency residual; Frequency-Averaged Mel
is then split along time into Coarse-Mel and a temporal residual.
\emph{Bottom to top}: The pipeline autoregressively generates Coarse-Mel,
then predicts the temporal and frequency residuals with non-autoregressive
tiers in dependency order. The inverse transform reconstructs the mel
spectrogram, which a vocoder converts to a waveform.}
  \label{fig:overview}
\end{figure*}

\subsection{Motivation for Coarse-Mel Autoregression}
\label{sec:method:motivation}

We analyze 2{,}000 utterances from LibriSpeech test-clean.
Adjacent mel frames have lag-one autocorrelation of $0.944$. A linear
predictor using only the preceding frame reduces target variance by
11.3\,dB (about 93\%); adding more than two past frames yields only
about 0.2\,dB of additional gain.

After the two-axis Haar split, Coarse-Mel accounts for 92.9\% of the
summed coefficient variance but only 25\% of the coefficients. The
temporal and frequency residuals account for 1.9\% and 5.2\%,
respectively. The temporal residual has little autocorrelation: using
its own past gives only 0.28\,dB of prediction gain. Its probe $R^2$
rises from about 0.01 with the same-position coarse coefficient to 0.57
when neighboring coarse frames are included. For the frequency residual,
same-frame coarse context gives a probe $R^2$ of about 0.42, with little
gain from adding temporal context. These measurements motivate excluding
both residuals from the autoregressive history. Section~\ref{sec:exp:ablations}
reports ablations on trained models.

\subsection{Two-Axis Haar Wavelet Transform}
\label{sec:method:haar}

A single-level Haar decomposition along time maps each adjacent frame pair
to a local average and half-difference:
\begin{equation}
  \begin{aligned}
    c_j &= \tfrac{1}{2}\!\left(x_{2j-1} + x_{2j}\right), \\
    d_j &= \tfrac{1}{2}\!\left(x_{2j-1} - x_{2j}\right),
  \end{aligned}
  \label{eq:haar}
\end{equation}
for $j=1, \dots, T/2$, with inverse $x_{2j-1}=c_j+d_j$ and $x_{2j}=c_j-d_j$. The same operation
along frequency pairs adjacent mel channels. VoxMel applies one split per
axis, first along frequency and then along time:
\begin{equation}
  \begin{aligned}
    (A,D^{(f)}) &= \mathrm{Haar}_{f}(X), \\
    (C,D^{(t)}) &= \mathrm{Haar}_{t}(A),
  \end{aligned}
\end{equation}
where $A$ is Frequency-Averaged Mel and $C$ is Coarse-Mel. The resulting
shapes are $D^{(f)}\in\mathbb{R}^{T\times M/2}$,
$D^{(t)}\in\mathbb{R}^{T/2\times M/2}$, and
$C\in\mathbb{R}^{T/2\times M/2}$. The inverse operations recover $X$
exactly, so the transform discards no coefficients. We use the averaging
form in Eq.~\ref{eq:haar}, so the coarse
coefficients remain on the mel-value scale and can use the same
normalization. This two-axis Haar decomposition is a critically sampled
discrete wavelet transform \citep{mallat1989wavelet}. Because each split
acts on adjacent pairs, each residual is local to its corresponding axis;
Section~\ref{sec:method:arch} describes how the residual tiers use coarse
context.

\subsection{Conditional Factorization}
\label{sec:method:factorization}

Let $H$ denote the fixed, invertible transform above, and let
$Z=H\,\mathrm{vec}(X)$ collect the Coarse-Mel grid $C$ and residual bands
$D^{(t)}$ and $D^{(f)}$. Here, $A$ is the Frequency-Averaged Mel intermediate reconstructed from $C$ and $D^{(t)}$. Write $c_i$ for the Coarse-Mel unit at position
$i$, and let $y$ denote the text and speaker conditions. The
change-of-variables formula gives the first equality below; the second
states VoxMel's modeling factorization:
\begin{equation}
  \begin{aligned}
    p_X(X\mid y) &= |\det H|\,p_Z(Z\mid y) \\
    &= |\det H|\Biggl[
      \underbrace{\prod_{i=1}^{T/2}p(c_i\mid c_{<i},y)}_{\substack{\text{autoregressive}\\[-2pt]\text{Coarse-Mel stream}}} \\
    &\qquad\cdot
      \underbrace{p(D^{(t)}\mid C)\,p(D^{(f)}\mid A)}_{\substack{\text{non-autoregressive}\\[-2pt]\text{residual prediction}}}
    \Biggr].
  \end{aligned}
  \label{eq:factorization}
\end{equation}
The Coarse-Mel units are sampled one at a time, conditioned on earlier
units and $y$. The residual tiers do not directly receive $y$. The temporal
residual is conditioned on Coarse-Mel; the frequency residual is conditioned
on $A$, which is reconstructed from Coarse-Mel and the temporal residual.
Thus, the sampling order follows the dependencies in the factorization:
Coarse-Mel first, then the temporal residual, and finally the frequency
residual. Within each residual tier, positions are predicted in parallel;
this parallelism is across positions, not across tiers. Neither residual
is fed back into the Coarse-Mel history. Since $H$ is
fixed and invertible, $|\det H|$ is a constant change-of-variables factor
and the transformed representation preserves all information needed to
reconstruct $X$.

\subsection{Architecture}
\label{sec:method:arch}

Figure~\ref{fig:overview} summarizes the data flow: the coarse stream is
generated autoregressively, the residual bands are added in coarse-to-fine
order, and the inverse Haar transform reconstructs the mel spectrogram.
The right side of the figure expands the residual tiers and the flow-matching
head used to sample each band.

\textbf{Autoregressive Coarse-Mel generation.} The transcript is tokenized,
and the text conditions are combined with projected Coarse-Mel
history as input to a decoder-only transformer. At each step, its hidden
state conditions a flow-matching head that samples the next Coarse-Mel unit;
a stop head predicts when to end the sequence. With frame-wise prediction,
there are $T/2$ steps, each of dimension $M/2$--half as many steps as
frame-level mel generation.
Inference applies separate guidance scales to text and speaker prompt conditions
(Section~\ref{sec:exp:guidance}).

\textbf{Residual generation and reconstruction.} The temporal-residual
tier uses a non-autoregressive transformer and a flow-matching head
conditioned on Coarse-Mel. Its prediction is combined with Coarse-Mel to
reconstruct Frequency-Averaged Mel, which conditions the
frequency-residual tier. This second tier predicts each frame
independently. Prediction is parallel within each tier but sequential
across tiers. The inverse Haar transform then reconstructs the full mel
spectrogram, which the vocoder converts to speech.

\begin{table*}[!t]
  \centering
  \caption{Objective results on LibriSpeech test-clean and Seed-TTS-eval
  test-en/test-zh for zero-shot voice cloning. WER and CER are reported
  in percent; lower is better. SIM denotes speaker similarity; higher is
  better.}
  \label{tab:main}
  \footnotesize
  \renewcommand{\arraystretch}{1.05}
  \setlength{\tabcolsep}{2.5pt}
  \begin{tabular}{@{}lcccccccc@{}}
    \toprule
    & & & \multicolumn{2}{c}{LibriSpeech test-clean} & \multicolumn{2}{c}{Seed-TTS test-en}
    & \multicolumn{2}{c}{Seed-TTS test-zh} \\
    \cmidrule(lr){4-5} \cmidrule(lr){6-7} \cmidrule(lr){8-9}
    System & Params & Hrs & WER $\downarrow$ & SIM $\uparrow$ & WER $\downarrow$ & SIM $\uparrow$ & CER $\downarrow$ & SIM $\uparrow$ \\
    \midrule
    \multicolumn{9}{l}{\emph{\textbf{Generation with Learned Tokens or Latents}}} \\
    CosyVoice \citep{du2024cosyvoice} & 0.3B & 170K & 3.59 & 0.660 & 3.39 & 0.629 & 3.63 & 0.723 \\
    CosyVoice 2 \citep{du2024cosyvoice2} & 0.5B & 167K & 2.46 & 0.679 & 2.57 & 0.652 & 1.45 & 0.748 \\
    CosyVoice 3 \citep{du2025cosyvoice3} & 1.1B & 1.0M & 1.99 & 0.697 & 2.02 & 0.718 & 1.19 & \VoxBestCell{0.780} \\
    Spark-TTS \citep{wang2025sparktts} & 0.5B & 102K & -- & -- & 1.98 & 0.584 & 1.20 & 0.672 \\
    MaskGCT \citep{wang2024maskgct} & 1.0B & 100K & 2.29 & 0.691 & 2.88 & 0.713 & 2.40 & 0.773 \\
    FireRedTTS 2 \citep{xie2025fireredtts2} & 1.5B & 1.4M & 4.32 & 0.642 & 1.95 & 0.665 & 1.14 & 0.736 \\
    Qwen3-TTS \citep{hu2026qwen3tts} & 1.7B & 5.0M & 2.35 & 0.703 & 1.66 & 0.714 & 1.22 & 0.770 \\
    IndexTTS 2 \citep{zhou2025indextts2} & 1.7B & 55K & 2.47 & 0.700 & 2.23 & 0.706 & 1.03 & 0.765 \\
    VibeVoice \citep{peng2025vibevoice} & 1.5B & -- & -- & -- & 3.04 & 0.689 & 1.16 & 0.744 \\
    MOSS-TTS \citep{gong2026mosstts} & 8.0B & -- & 1.98 & 0.677 & 1.84 & 0.709 & 1.37 & 0.770 \\
    MOSS-TTS-Nano \citep{gong2026mosstts} & 0.1B & -- & 4.10 & 0.484 & 4.62 & 0.499 & 3.13 & 0.643 \\
    DiTAR \citep{jia2025ditar} & 0.6B & 280K & 2.39 & 0.670 & 1.69 & 0.735 & 1.02 & 0.753 \\
    VoxCPM \citep{zhou2025voxcpm} & 0.7B & 1.8M & 1.74 & 0.717 & 1.85 & 0.729 & 0.93 & 0.772 \\
    OmniVoice-Emilia \citep{omnivoice} & 0.8B & 100K & \VoxBestCell{1.57} & 0.697 & 1.75 & 0.717 & \VoxBestCell{0.89} & 0.765 \\
    \midrule
    \multicolumn{9}{l}{\emph{\textbf{Non-Autoregressive with Mel}}} \\
    E2-TTS (32 NFE) \citep{eskimez2024e2} & 0.3B & 100K & 2.95 & 0.690 & 2.32 & 0.706 & 1.91 & 0.713 \\
    F5-TTS (32 NFE) \citep{chen2024f5} & 0.3B & 100K & 2.42 & 0.660 & 1.85 & 0.664 & 1.53 & 0.750 \\
    ZipVoice (16 NFE) \citep{zhu2025zipvoice} & 0.1B & 100K & 2.05 & 0.674 & 1.70 & 0.697 & 1.40 & 0.751 \\
    \midrule
    \multicolumn{9}{l}{\emph{\textbf{Autoregressive with Mel }}} \\
    MELLE \citep{meng2024melle} & -- & 50K & 2.10 & 0.625 & -- & -- & -- & -- \\
    MELLE-R2 \citep{meng2024melle} & -- & 50K & 2.14 & 0.608 & -- & -- & -- & -- \\
    \textbf{VoxMel} (Ours, 32 NFE) & 0.6B & 100K & 1.98 & 0.721 & 1.51 & 0.738 & 1.29 & 0.755 \\
    \; + GRPO & 0.6B & 100K & 1.95 & \VoxBestCell{0.736} & \VoxBestCell{1.15} & \VoxBestCell{0.753} & 1.17 & 0.776 \\
    \bottomrule
  \end{tabular}

\end{table*}

\textbf{Flow-matching head.} As shown in the lower-right inset, each head
takes a noisy input together with the flow time and its conditioning signal
(the transformer state for Coarse-Mel, or coarse-level context for a
residual). An MLP predicts the flow, and an ODE solver integrates it to
produce a Coarse-Mel unit or residual sample.

\textbf{Training.} Each flow-matching head uses the same conditional
flow-matching objective \citep{lipman2022flow}:
\begin{equation}
  \begin{aligned}
    \mathcal{L}_{\mathrm{CFM}}
    &= \mathbb{E}_{(a,g),s,\epsilon}\Bigl[
      \bigl\|v_\phi(z_s,s,g) - (a-\epsilon)\bigr\|_2^2
      \Bigr], \\
    z_s &= (1-s)\epsilon+s a.
  \end{aligned}
  \label{eq:cfm}
\end{equation}
Here, $a$ is the target from the stream being modeled and $g$ is its
conditioning signal, computed or provided for the selected stream from
its training example. The flow time $s$ is sampled uniformly from $[0,1]$, and
$\epsilon$ is standard Gaussian noise with the same shape as $a$. Thus,
$z_s$ interpolates linearly from noise to target, and $a-\epsilon$ is the
target velocity. The head $v_\phi$ predicts this velocity, with $\phi$
denoting that head's parameters. For Coarse-Mel, $a=c_i$ and
$g=h_i=f_\theta(c_{<i},y)$, where $h_i$ is computed from the
teacher-forced coarse prefix and the text and speaker conditions $y$, and
$\theta$ denotes the backbone parameters. For the temporal and
frequency residual heads, $(a,g)=(D^{(t)},C)$ and
$(D^{(f)},A)$, respectively. Each tier predicts all positions without
autoregressive dependencies, conditioned on its context. During training of the
autoregressive Coarse-Mel stream, we use condition
dropout: examples have both text and speaker prompt, text dropped with
the speaker prompt retained, or both conditions dropped. These are the
three condition states used for two-scale guidance at inference; the
full schedule is given in Section~\ref{sec:exp:guidance}. Full training details
are in Section~\ref{sec:exp:setup}.

\subsection{GRPO with a Frozen Flow Head}
\label{sec:method:grpo}

We apply Group Relative Policy Optimization (GRPO)
\citep{shao2024deepseekmath} after supervised training to optimize the
autoregressive Coarse-Mel policy. Motivated by prior work that
freezing the diffusion head can stabilize RL in autoregressive--diffusion
hybrids \citep{ma2026margrpo}, we adopt this strategy for VoxMel by
updating only the backbone and keeping the Coarse-Mel flow head, both
residual tiers, and the vocoder frozen. This keeps the mapping from the
backbone's predictions to audio fixed throughout RL, isolating policy
updates from changes to the acoustic decoder. We evaluate this choice in
Section~\ref{sec:exp:grpo}.

For each text-and-speaker prompt, we sample a group of eight ($G=8$)
outputs using the deployment two-scale classifier-free guidance
scales, rather than sampling from the unguided conditional model.
Each output receives a scalar reward. GRPO normalizes rewards within
each prompt's group to estimate relative advantages: an output is favored
when it scores better than the other outputs for the same prompt. A KL
penalty to the supervised checkpoint limits how far the updated policy
can drift. The reward design and rollout and optimization settings are
detailed in Section~\ref{sec:exp:grpo}.

\subsection{Guided Flow-Head Distillation}
\label{sec:method:distill}

Sampling cost remains in the Coarse-Mel autoregressive loop. Each
coarse frame requires an ODE solve of the flow head. With two-scale
guidance, each solver step requires three forward passes, corresponding
to the three condition states described in
Section~\ref{sec:method:arch} (full dropout schedule in
Section~\ref{sec:exp:guidance}). Thus, $S$ solver steps require $3S$ head
evaluations per coarse frame.

The guidance factor of three is removable because guidance is a
\emph{deterministic} function of the conditional and dropped-condition
velocities. We distill a few-step student head that absorbs guidance
into its own weights: it matches a $4$-step guided teacher trajectory
while taking the guidance scales $(\omega_{\text{text}},
\omega_{\text{spk}})$ as explicit inputs, sampled during training from
ranges spanning the useful operating region. At inference the student
uses the requested scales and one conditional forward pass
per solver step, thereby avoiding separate dropped-condition passes while keeping the scales tunable
without retraining. Distillation starts from the GRPO checkpoint and
updates only the head, with the backbone and tiers fixed. We draw a
fraction of training batches from the student's own rollouts, so the
head trains on the conditioning it will meet at inference rather than on
ground-truth prefixes alone; without this the distilled head degrades
severely (Section~\ref{sec:exp:efficiency}).

% ======================================================================
\section{Experiments}
\label{sec:experiments}

We first report benchmark and model-scaling results, then evaluate GRPO
post-training and sampling efficiency, and finally test transfer to a
DiTAR-style backbone. Decomposition ablations and their mechanism probes
are in Section~\ref{sec:exp:ablations}.

\subsection{Experimental Setup}
\label{sec:exp:setup}

We evaluate three VoxMel sizes (0.15B, 0.3B, and 0.6B), which share the
same wavelet target and training data but vary in backbone and tier
configuration. Unless stated otherwise, results are for VoxMel-0.6B.

\paragraph{Architecture and model sizes.} The backbone is initialized
from Qwen2.5-0.5B and used as a feature extractor. Table~\ref{tab:params}
reports the 0.6B module counts; Table~\ref{tab:sizes} lists all three
sizes. The temporal residual tier uses bidirectional attention, while
the frequency tier processes each frame independently.

\begin{table*}[t]
  \centering
  \caption{VoxMel-0.6B module configurations and parameter counts.}
  \label{tab:params}
  \small
  \begin{tabular}{@{}llr@{}}
    \toprule
    Module & Configuration & Params \\
    \midrule
    Backbone & 24 layers, 896 wide, 4864 FFN, initialized from Qwen2.5-0.5B & 494.0M \\
    Coarse head & flow matching, 6 layers, 1024 wide & 14.8M \\
    Temporal residual & 6 bidirectional attention layers, 6-layer flow head, 1024 wide & 90.6M \\
    Frequency residual & 6 frame-wise MLP layers, 6-layer flow head, 1024 wide & 90.6M \\
    Adapters & coarse-to-backbone projection, special embeddings, stop head & 0.1M \\
    \midrule
    Total & & 690.1M \\
    \midrule
    Vocoder & Vocos, 8 layers, 512 wide (frozen, not counted above) & 13.6M \\
    \bottomrule
  \end{tabular}
\end{table*}

\begin{table}[t]
  \centering
  \caption{The three VoxMel model sizes.}
  \label{tab:sizes}
  \small
  \setlength{\tabcolsep}{4pt}
  \begin{tabular}{@{}lrrr@{}}
    \toprule
    Parameters& 0.15B & 0.3B & 0.6B \\
    \midrule
    \multicolumn{4}{c}{\emph{Configuration}} \\
    Backbone layers & 16 & 20 & 24 \\
    Backbone width & 512 & 768 & 896 \\
    Backbone FFN & 1024 & 1792 & 4864 \\
    Residual tier width & 512 & 640 & 1024 \\
    Blocks per tier & 4 & 5 & 6 \\
    Flow head layers & 4 & 5 & 6 \\
    \midrule
    \multicolumn{4}{c}{\emph{Parameters}} \\
    Backbone & 119.8M & 246.5M & 494.0M \\
    Coarse head & 2.7M & 5.2M & 14.8M \\
    Temporal residual & 15.4M & 30.2M & 90.6M \\
    Frequency residual & 15.4M & 30.2M & 90.6M \\
    Adapters & 0.1M & 0.1M & 0.1M \\
    \midrule
    Total & 153.4M & 312.2M & 690.1M \\
    \bottomrule
  \end{tabular}
\end{table}

\paragraph{Data and acoustic front-end.} We train on 100K hours of
Emilia~\citep{he2024emilia}, discarding utterances shorter than $0.3$\,s
or longer than $30$\,s. Batches use a budget of $24{,}000$ frames and a
maximum sequence length of $4096$ tokens. We use the Qwen2.5 tokenizer.
The target is a 100-channel log-mel spectrogram at $24$\,kHz, computed
with an FFT/window size of $1024$ and hop of $256$ samples
($93.75$ frames/s); a frozen Vocos vocoder reconstructs 24-kHz waveforms.

\paragraph{Optimization and inference.} We use AdamW with learning rate
$2\times10^{-4}$, cosine decay, and $10{,}000$ warmup steps. The
supervised 0.6B checkpoint is taken after $900{,}000$ steps. Gradients
are clipped at global norm $2$; training uses bfloat16 without gradient
accumulation and an exponential moving average for evaluation. Sampling
uses an Euler solver with 32 uniform steps for the Coarse-Mel head and
each residual tier.

\paragraph{Evaluation.} We report WER, CER, and speaker-embedding cosine
similarity (SIM), averaged over utterances, on LibriSpeech test-clean
\citep{chen2024f5} and Seed-TTS-eval test-en (1{,}088 utterances) and
test-zh (2{,}020 utterances) \citep{anastassiou2024seed}. English transcription uses faster-whisper large-v3, Chinese
uses FunASR Paraformer-zh, and speaker similarity uses WavLM-large
embeddings. Inference uses guidance scales
$(\gamma_{\text{text}},\gamma_{\text{spk}})=(3.5,3.0)$, selected in
the guidance sweep above.

\subsection{Classifier-Free Guidance}
\label{sec:exp:guidance}

We use nested text-first condition dropout. Here, $T$ and $S$ denote
the text and speaker prompt conditions, respectively. For each example,
we independently sample $r_{\text{text}}\sim\mathrm{Bern}(0.3)$ and
$r_{\text{spk}}\sim\mathrm{Bern}(0.5)$, then set
\begin{equation}
  d_{\text{text}}=r_{\text{text}},\qquad
  d_{\text{spk}}=r_{\text{text}}\wedge r_{\text{spk}},
  \label{eq:nested-drop}
\end{equation}
where $d=1$ denotes a dropped condition. The resulting states
$(T,S)$, $(\emptyset,S)$, and $(\emptyset,\emptyset)$ occur with
probabilities $0.70$, $0.15$, and $0.15$; $(T,\emptyset)$ is not used.
Inference uses the matching two-scale guidance path:
\begin{equation}
  \begin{aligned}
    v ={}& v(\emptyset,\emptyset)
      + \gamma_{\text{text}}\,[v(T,S)-v(\emptyset,S)] \\
      &+ \gamma_{\text{spk}}\,[v(\emptyset,S)-v(\emptyset,\emptyset)],
  \end{aligned}
  \label{eq:two-scale-cfg}
\end{equation}

When speaker guidance is much stronger than text guidance, intelligibility
degrades. We select $(3.5,3.0)$ as a balanced operating point: it lies
near the CER--SIM trade-off frontier while CER remains stable.

\subsection{Benchmark Results}
\label{sec:exp:main}

On Seed-TTS-eval test-en, supervised VoxMel-0.6B obtains
1.51\% WER and 0.738 SIM, compared with 1.69\% and 0.735 for DiTAR
and 1.70\% and 0.697 for ZipVoice (Table~\ref{tab:main}).  On test-zh, VoxMel obtains 1.29\% CER and 0.755 SIM. Several
systems have lower CER, while the GRPO variant improves VoxMel's results
to 1.15\% WER / 0.753 SIM on test-en and 1.17\% CER / 0.776 SIM on
test-zh.

On LibriSpeech test-clean, VoxMel-0.6B obtains 1.98\% WER and 0.721
SIM; GRPO changes these to 1.95\% and 0.736. The GRPO model has the
highest SIM in the table, while its WER remains above OmniVoice-Emilia
(1.57\%) and VoxCPM (1.74\%).

We also compared VoxMel with four baselines in a five-point listening
test. VoxMel's naturalness/similarity scores are 3.89/3.97 in Chinese
and 3.96/4.11 in English. The highest scores vary by system and metric;
Table~\ref{tab:mos} reports means and 95\% confidence intervals.

\newcommand{\sd}[1]{\scriptsize$\pm$#1}
\begin{table}[t]
  \centering
  \caption{Five-point listening-test ratings for all listed systems.
  N-MOS measures naturalness; S-MOS measures speaker similarity. Values
  are means with 95\% confidence intervals.}
  \label{tab:mos}
  \footnotesize
  \setlength{\tabcolsep}{2pt}
  \renewcommand{\arraystretch}{1.1}
  \begin{tabular}{@{}lcccc@{}}
    \toprule
    & \multicolumn{2}{c}{ZH} & \multicolumn{2}{c}{EN} \\
    \cmidrule(lr){2-3} \cmidrule(lr){4-5}
    System & N-MOS & S-MOS & N-MOS & S-MOS \\
    \midrule
    MaskGCT & 3.36\sd{0.11} & 3.89\sd{0.11} & 3.83\sd{0.11} & 4.00\sd{0.10} \\
    CosyVoice 2 & 3.56\sd{0.12} & 3.93\sd{0.10} & 4.01\sd{0.09} & 3.95\sd{0.10} \\
    IndexTTS 2 & 4.19\sd{0.09} & 4.01\sd{0.09} & 4.06\sd{0.10} & 4.16\sd{0.09} \\
    VoxCPM & 3.83\sd{0.12} & 3.98\sd{0.11} & 3.99\sd{0.10} & 4.17\sd{0.09} \\
    \midrule
    VoxMel (ours) & 3.89\sd{0.10} & 3.97\sd{0.10} & 3.96\sd{0.09} & 4.11\sd{0.09} \\
    \bottomrule
  \end{tabular}
\end{table}

\paragraph{Listening protocol.} Twenty native speakers per language
rated naturalness and speaker similarity on separate five-point scales.
Naturalness raters heard a synthesized clip with its text; similarity
raters compared reference and synthesized clips while judging voice
identity independently of content and recording quality.

\subsection{Decomposition Ablations}
\label{sec:exp:ablations}

All ablation variants use the VoxMel-0.15B configuration. For
Seed-TTS-eval, we report results on test-en only.

We compare full-band mel (none), frequency residual excluded ($f$),
temporal residual excluded ($t$), and both residuals excluded ($t{+}f$).
Excluded residuals are generated by separate non-autoregressive tiers.
The text-reliance probe measures the loss increase when correct text is
removed at a fixed acoustic prefix. The exposure gap $G$ compares
self-sampled and teacher-forced losses after rollouts re-anchored at
ground-truth prefixes.

\begin{table}[t]
  \centering
  \caption{Frame-level decomposition ablations using VoxMel-0.15B on
  Seed-TTS-eval test-en. WER is reported in percent (lower is better);
  SIM denotes speaker similarity (higher is better).}
  \label{tab:ablation}
  \small
  \setlength{\tabcolsep}{3pt}
  \begin{tabular}{@{}llccccc@{}}
    \toprule
    & & & \multicolumn{2}{c}{probe} & \multicolumn{2}{c}{end-to-end} \\
    \cmidrule(lr){4-5} \cmidrule(lr){6-7}
    Split & Recursion & Hz & td\% & $G$ & WER & SIM \\
    \midrule
    none    & full-band mel & 94 & 1.10 & 16.1 & 30.35 & 0.493 \\
    $f$     & freq.\ out    & 94 & 2.02 & 18.9 &  6.64 & 0.621 \\
    $t$     & temp.\ out    & 47 & 3.07 &  5.4 & 11.41 & 0.552 \\
    $t{+}f$ & both out      & 47 & 4.53 &  6.7 &  1.90 & 0.667 \\
    \bottomrule
  \end{tabular}
\end{table}

Removing residuals from the autoregressive history increases measured
text reliance from 1.10\% to 4.53\%. The full-band model has the highest
WER, while the fully split model has the lowest WER and highest SIM. the probes therefore
measure different effects of the decomposition.

\subsection{Model Scaling}
\label{sec:exp:scaling}

From 0.15B to 0.6B, English WER falls from 1.90\% to 1.51\%, and
speaker similarity rises from 0.667 to 0.738 on test-en and from 0.730
to 0.755 on test-zh. Chinese CER is lowest at 1.17\% for the 0.3B
model; it is 1.44\% for the 0.15B model and 1.29\% for the 0.6B model
(Table~\ref{tab:scaling}).

\begin{table}[t]
  \centering
  \caption{VoxMel scaling results on Seed-TTS-eval.}
  \label{tab:scaling}
  \small
  \setlength{\tabcolsep}{3pt}
  \begin{tabular}{@{}lcccc@{}}
    \toprule
    & \multicolumn{2}{c}{test-en} & \multicolumn{2}{c}{test-zh} \\
    \cmidrule(lr){2-3} \cmidrule(lr){4-5}
    Model & WER $\downarrow$ & SIM $\uparrow$ & CER $\downarrow$ & SIM $\uparrow$ \\
    \midrule
    VoxMel-0.15B & 1.90 & 0.667 & 1.44 & 0.730 \\
    VoxMel-0.3B & 1.54 & 0.715 & 1.17 & 0.748 \\
    VoxMel-0.6B & 1.51 & 0.738 & 1.29 & 0.755 \\
    \bottomrule
  \end{tabular}
\end{table}

\subsection{GRPO Post-Training}
\label{sec:exp:grpo}

We compare two GRPO variants initialized from the same supervised
0.6B checkpoint. Both update the backbone and use identical guided
rollouts, rewards, and optimization settings. The trainable-head variant
also updates the Coarse-Mel flow head, whereas the frozen-head variant
keeps it fixed; the residual tiers remain frozen in both. We sample eight guided
rollouts per prompt and combine speaker similarity with ASR-based
transcription accuracy using equal reward weights. Advantages are
normalized within each group, with a KL penalty ($\beta=0.01$) to the
supervised checkpoint. Rollouts use 10 Euler steps, injected noise
$\sigma=0.15$, and learning rate $2\times10^{-6}$. GRPO uses 10{,}000 Emilia examples for 1{,}000
updates.

\begin{table}[t]
  \centering
  \caption{Trainable- versus frozen-head GRPO on Seed-TTS-eval, the pretrained row is the supervised baseline. Lower
  WER/CER and higher SIM are better.}
  \label{tab:fixhead}
  \small
  \setlength{\tabcolsep}{2.5pt}
  \begin{tabular}{@{}l
      S[table-format=4.0]
      S[table-format=1.2] S[table-format=1.3]
      S[table-format=1.2] S[table-format=1.3]@{}}
    \toprule
    & & \multicolumn{2}{c}{test-en} & \multicolumn{2}{c}{test-zh} \\
    \cmidrule(lr){3-4} \cmidrule(lr){5-6}
    Training setup & {RL steps} & {WER $\downarrow$} & {SIM $\uparrow$} & {CER $\downarrow$} & {SIM $\uparrow$} \\
    \midrule
    Pretrained & {--} &  1.51 & 0.738 & 1.29 & 0.755 \\
    \midrule
    GRPO: trainable & 1000 & 3.21 & 0.715 & 4.27 & 0.733 \\
    GRPO: frozen & 1000 & \bfseries 1.15 & \bfseries 0.753 & \bfseries 1.17 & \bfseries 0.776 \\
    \bottomrule
  \end{tabular}
\end{table}

After 1{,}000 GRPO steps, the frozen-head variant improves all four
Seed-TTS-eval metrics over the supervised baseline: test-en WER falls
from 1.51\% to 1.15\% and SIM rises from 0.738 to 0.753; test-zh CER
falls from 1.29\% to 1.17\% and SIM rises from 0.755 to 0.776. The
trainable-head variant performs worse than both the frozen-head variant
and the supervised baseline on all four metrics (Table~\ref{tab:fixhead}).

\subsection{Sampling Efficiency}
\label{sec:exp:efficiency}

\paragraph{Solver-Step Sweep.}
We measure the quality--compute trade-off by varying the number of
flow-solver steps for the supervised 0.6B model. For each setting, both
residual tiers use the same number of solver steps as the autoregressive
Coarse-Mel flow head. At each step, two-scale guidance requires three
forward passes through the Coarse-Mel head, so its number of function
evaluations (NFE) is three times the solver-step count. The NFE values count only Coarse-Mel head evaluations, not those of the residual
tiers.

\begin{table}[t]
  \centering
  \caption{Solver-step sweep for supervised VoxMel-0.6B on
  Seed-TTS-eval. NFE is head evaluations per Coarse-Mel frame; lower
  NFE/WER/CER and higher SIM are better.}
  \label{tab:steps}
  \small
  \setlength{\tabcolsep}{4pt}
  \begin{tabular}{@{}S[table-format=2.0] S[table-format=2.0]
      S[table-format=2.2] S[table-format=1.3]
      S[table-format=2.2] S[table-format=1.3]@{}}
    \toprule
    & & \multicolumn{2}{c}{test-en} & \multicolumn{2}{c}{test-zh} \\
    \cmidrule(lr){3-4} \cmidrule(lr){5-6}
    {Steps} & {NFE $\downarrow$} & {WER $\downarrow$} & {SIM $\uparrow$} & {CER $\downarrow$} & {SIM $\uparrow$} \\
    \midrule
    1  & 3  & 12.21 & 0.637 & 10.32 & 0.644 \\
    2  & 6  & 9.65  & 0.690 & 5.46  & 0.716 \\
    4  & 12 & 3.10  & 0.726 & 2.61  & 0.751 \\
    8  & 24 & \bfseries 1.38  & \bfseries 0.741 & 1.39 & \bfseries 0.761 \\
    32 & 96 & 1.51 & 0.738 & \bfseries 1.29 & 0.755 \\
    \bottomrule
  \end{tabular}
\end{table}

Reducing the solver from 32 to 8 steps lowers Coarse-Mel head NFE from
96 to 24. On test-en, WER and SIM improve from 1.51\% and 0.738 to
1.38\% and 0.741. On test-zh, SIM rises from 0.755 to 0.761, while CER
increases from 1.29\% to 1.39\%. At four or fewer steps, both error
rates rise substantially.
\paragraph{Few-Step Guided Distillation.}
Following Section~\ref{sec:method:distill}, we distill the GRPO
checkpoint's Coarse-Mel flow head to match the trajectory of a guided,
4-step teacher. The student takes the guidance scales as inputs, sampled
during training from $\omega_{\text{text}} \in [3,4]$ and
$\omega_{\text{spk}} \in [2,4]$. We keep the backbone and residual tiers
fixed, use a learning rate of $5\times10^{-5}$, and draw $30\%$ of
training batches from 48-frame student rollouts. At inference, the
student uses one pure-conditional forward pass per solver step, with the
desired guidance scales supplied as inputs. We evaluate it at
$(\omega_{\text{text}},\omega_{\text{spk}}) = (3.5,3)$.

Table~\ref{tab:distill} compares the guided GRPO checkpoint with the
distilled student at one, two, and four inference steps, using the same
guidance scales.

\begin{table}[t]
  \centering
  \caption{Few-step guided distillation on Seed-TTS-eval. NFE counts
  head evaluations per Coarse-Mel frame.}
  \label{tab:distill}
  \scriptsize
  \setlength{\tabcolsep}{3pt}
  \begin{tabular}{@{}l
      S[table-format=2.0]
      S[table-format=1.2] S[table-format=1.3]
      S[table-format=1.2] S[table-format=1.3]@{}}
    \toprule
    & & \multicolumn{2}{c}{test-en} & \multicolumn{2}{c}{test-zh} \\
    \cmidrule(lr){3-4} \cmidrule(lr){5-6}
    Model & {NFE $\downarrow$} & {WER $\downarrow$} & {SIM $\uparrow$} & {CER $\downarrow$} & {SIM $\uparrow$} \\
    \midrule
    GRPO, guided head & 96 & 1.15 & 0.752 & 1.17 & 0.773 \\
    distilled, 4 steps & 4 & 1.49 & 0.693 & 1.20 & 0.747 \\
    distilled, 2 steps & 2 & 1.75 & 0.693 & 1.57 & 0.744 \\
    distilled, 1 step & 1 & 3.13 & 0.681 & 3.39 & 0.737 \\
    \bottomrule
  \end{tabular}
\end{table}

With four inference steps, the student reduces Coarse-Mel head
evaluations from 96 to 4 per frame ($24\times$). It combines four solver
steps with one forward pass per step by absorbing guidance into the head.
Compared with the guided GRPO checkpoint, test-en WER rises from 1.15\%
to 1.49\% and test-zh CER from 1.17\% to 1.20\%. SIM falls from 0.752
to 0.693 on test-en and from 0.773 to 0.747 on test-zh. The one- and
two-step students have higher error rates still (Table~\ref{tab:distill}).

\subsection{Transfer to a DiTAR-Style Backbone}
\label{sec:exp:transfer}

We assess whether wavelet factorization extends beyond the VoxMel
backbone by adapting a DiTAR-style patch-autoregressive model to the
wavelet-factorized mel target. We compare this variant with models using
full-band mel and VoxCPM VAE latents. The DiTAR-style model uses
\texttt{patch\_size}=4 and \texttt{history\_path\_size}=4.

\begin{table}[t]
  \centering
  \caption{Autoregressive target comparison with a DiTAR-style
  patch-autoregressive backbone on Seed-TTS-eval.}
  \label{tab:ditar}
  \small
  \setlength{\tabcolsep}{3pt}
  \renewcommand{\arraystretch}{1.1}
  \begin{tabular}{@{}l
      S[table-format=2.2] S[table-format=1.3]
      S[table-format=2.2] S[table-format=1.3]@{}}
    \toprule
    & \multicolumn{2}{c}{test-en} & \multicolumn{2}{c}{test-zh} \\
    \cmidrule(lr){2-3} \cmidrule(lr){4-5}
    Autoregressive target & {WER $\downarrow$} & {SIM $\uparrow$} & {CER $\downarrow$} & {SIM $\uparrow$} \\
    \midrule
    full-band mel & 36.97 & 0.454 & 49.99 & 0.379 \\
    VoxCPM VAE latents & 1.77 & 0.644 & 1.76 & 0.707 \\
    wavelet-factorized mel & 1.89 & 0.660 & 1.74 & 0.725 \\
    \bottomrule
  \end{tabular}
\end{table}

The wavelet-based variant obtains 1.89\% WER / 0.660 SIM on test-en
and 1.74\% CER / 0.725 SIM on test-zh, compared with 1.77\% / 0.644 and
1.76\% / 0.707 for the VAE-latent variant. The configurations are not
fully matched, so these results do not establish that either target is
better. They show that a DiTAR-style backbone can be trained on the fixed
wavelet representation without a learned tokenizer.

% ======================================================================
\section{Related Work}
\label{sec:related}

\paragraph{Mel as a synthesis target.}
Before language-model-based speech generation, mel spectrograms were a
standard acoustic target for TTS systems such as Tacotron~2,
Transformer-TTS, and FastSpeech~2. These systems predict mel features
and rely on a separate vocoder to synthesize waveforms
\citep{shen2018tacotron2,li2019transformertts,ren2020fastspeech2}.
VoxMel retains mel's interpretable, tokenizer-free representation, but
changes how it is used autoregressively: only the coarse wavelet
component, rather than the full frame, enters the history.

\paragraph{Learned speech representations.}
Many zero-shot TTS systems instead generate learned audio
representations, including discrete codec tokens and continuous
latents. This line of work spans AudioLM and VALL-E
\citep{borsos2023audiolm,wang2023valle}, as well as the CosyVoice,
IndexTTS, and FireRedTTS families
\citep{du2024cosyvoice,du2024cosyvoice2,du2025cosyvoice3,deng2025indextts,zhou2025indextts2,xie2025fireredtts2}.
The representations range from residual and scalar quantization
\citep{defossez2022encodec,kumar2023dac,mentzer2024fsq} to decoupled
token streams \citep{wang2025sparktts} and end-to-end learned latents
\citep{zhou2025voxcpm}. In contrast to these learned
representations, VoxMel applies a fixed, invertible coarse--residual
transform directly to mel features.

\paragraph{Continuous generation and history robustness.}
MAR provides a related precedent in image generation, using a per-step
diffusion head to generate continuous image patches rather than discrete
tokens \citep{li2024mar}. Speech models employ related generative heads,
but differ from MAR in both architecture and target representation.
MELLE and FELLE autoregress directly over mel-spectrogram frames, using
variational sampling and flow matching, respectively. ARDiT, DiTAR, and
CLEAR instead model learned continuous acoustic latents; these systems
use learned autoencoder-based representations, including VAEs
\citep{meng2024melle,wang2025felle,liu2024ardit,jia2025ditar,wu2025clear}.
VibeVoice also uses low-rate learned acoustic latents
\citep{peng2025vibevoice}. These models differ in what enters the
autoregressive history. DiTAR
feeds generated latent patches into later steps, while MELLE and FELLE
feed back generated mel frames; FELLE's within-step refinement does not
change that feedback. VoxMel instead keeps its residual bands out of the
history. Non-autoregressive flow and diffusion models avoid this
recursion altogether
\citep{le2023voicebox,shen2023naturalspeech2,guan2024reflow,eskimez2024e2,chen2024f5,zhu2025zipvoice}.

\paragraph{Multiscale factorization.}
Prior hierarchical speech models learn their coarse-to-fine structure:
HALL-E and MARS6 use layered codecs at different temporal rates, while
HAM-TTS adds a learned acoustic hierarchy over tokens
\citep{nishimura2025halle,baas2025mars6,wang2024hamtts}. VoxMel instead
obtains its levels from a fixed Haar transform. This linear split is
exactly invertible and adds no trainable parameters; we use it to keep
the coarse component in the autoregressive history and generate the
residuals separately.

% ======================================================================
\section{Conclusion}
\label{sec:conclusion}

VoxMel offers a learned-tokenizer-free approach to autoregressive mel-spectrogram modeling. A fixed, exactly invertible two-axis Haar transform
separates a coarse sequence from temporal and frequency residuals. The
model feeds only the coarse sequence into its autoregressive history and
predicts the residuals in separate non-autoregressive tiers, preserving
the full mel representation without autoregressive residual feedback.

Our experiments show that this factorization encourages greater reliance
on text than full-band mel autoregression while supporting effective
zero-shot synthesis. Frozen-head GRPO further improves synthesis quality.
Few-step distillation reduces sampling cost while retaining transcription
accuracy close to the supervised model and competitive speaker similarity.
Together, these results show that separating coarse structure from
residual detail is a practical way to model mel-spectrograms
autoregressively.

\section*{Limitations}
\paragraph{Depth and choice of transform.} We use one Haar split per
axis, motivated by the measured first-order temporal redundancy. We do not evaluate alternative depths,
other wavelet families in this work.

\paragraph{Languages and speech styles.} We evaluate English and
Chinese on benchmarks of single-speaker utterances a few seconds long.
Tonal languages beyond Chinese, richer morphology, languages requiring
specialized text front ends, long-form synthesis, multi-speaker dialogue,
expressive and paralinguistic control, singing, and code-switching are
all outside this evaluation.

\paragraph{Other audio domains.} Our evaluation is limited to speech.
Although the Haar transform itself is not speech-specific, we have not
tested whether its subband structure or the proposed autoregressive
factorization is effective for music, general audio, or sound effects.
Evaluating these domains, as well as alternative time--frequency
representations, remains future work.

\section*{Ethics Statement}
VoxMel supports zero-shot voice cloning from reference speech, which may
enable impersonation, non-consensual voice synthesis, or attempts to
spoof speaker-verification systems. Any release should document training
data provenance and consent, define acceptable uses, and consider
safeguards such as output watermarking and evaluated synthetic-speech
detection.

\clearpage

\bibliography{voxmel_references}

% =====================================================================
\clearpage

\end{document}